\documentclass[conference]{IEEEtran}
\IEEEoverridecommandlockouts
\usepackage{cite}
\usepackage{amsmath,amssymb,amsfonts}
\usepackage{algorithmic}
\usepackage{graphicx}
\usepackage{textcomp}
\usepackage{multirow}
\usepackage{url}
\usepackage[hidelinks]{hyperref}
\usepackage[table]{xcolor}
\usepackage{array}
\usepackage{enumitem}
\usepackage{pgf}
\usepackage{collcell}

\newcommand*{\MinNumber}{0.39}%
\newcommand*{\MaxNumber}{0.67}%
\newcommand{\ApplyGradient}[1]{%
  \pgfmathsetmacro{\PercentColor}{75.0*(#1-\MinNumber)/(\MaxNumber-\MinNumber)}%
  \edef\x{\noexpand\cellcolor{green!\PercentColor}}\x\textcolor{black}{#1}%
}
\newcolumntype{E}{>{\collectcell\ApplyGradient}{r}<{\endcollectcell}}

\newcommand*{\MinNumberSec}{0.84}%
\newcommand*{\MaxNumberSec}{1}%
\newcommand{\ApplyGradientSec}[1]{%
  \pgfmathsetmacro{\PercentColor}{75.0*(#1-\MinNumberSec)/(\MaxNumberSec-\MinNumberSec)}%
  \edef\x{\noexpand\cellcolor{green!\PercentColor}}\x\textcolor{black}{#1}%
}
\newcolumntype{F}{>{\collectcell\ApplyGradientSec}{r}<{\endcollectcell}}

\newcolumntype{R}[1]{>{\raggedleft\let\newline\\\arraybackslash\hspace{0pt}}m{#1}}
\newcolumntype{L}[1]{>{\raggedright\let\newline\\\arraybackslash\hspace{0pt}}m{#1}}
\def\BibTeX{{\rm B\kern-.05em{\sc i\kern-.025em b}\kern-.08em
    T\kern-.1667em\lower.7ex\hbox{E}\kern-.125emX}}
\begin{document}

\title{Towards Extracting Software Requirements from App Reviews using Seq2seq Framework}

\author{
    \IEEEauthorblockN{Aakash Sorathiya, Gouri Ginde}
    \IEEEauthorblockA{\textit{Department of Electrical and Software Engineering} \\
        \textit{University of Calgary}\\
        Calgary, Canada \\
        \{aakash.sorathiya, gouri.ginde\}@ucalgary.ca}
}

\maketitle

\begin{abstract}
Mobile app reviews are a large-scale data source for software improvements. A key task in this context is effectively extracting requirements from app reviews to analyze the users' needs and support the software's evolution. Recent studies show that existing methods fail at this task since app reviews usually contain informal language, grammatical and spelling errors, and a large amount of irrelevant information that might not have direct practical value for developers. To address this, we propose a novel reformulation of requirements extraction as a Named Entity Recognition (NER) task based on the sequence-to-sequence (Seq2seq) generation approach. With this aim, we propose a Seq2seq framework, incorporating a BiLSTM encoder and an LSTM decoder, enhanced with a self-attention mechanism, GloVe embeddings, and a CRF model. We evaluated our framework on two datasets: a manually annotated set of 1,000 reviews (Dataset 1) and a crowdsourced set of 23,816 reviews (Dataset 2). The quantitative evaluation of our framework showed that it outperformed existing state-of-the-art methods with an F1 score of 0.96 on Dataset 2, and achieved comparable performance on Dataset 1 with an F1 score of 0.47.
\end{abstract}

\begin{IEEEkeywords}
Requirements extraction, app reviews, sequence-to-sequence modeling, Named Entity Recognition (NER)
\end{IEEEkeywords}

\section{Introduction}
\textbf{Context. }The rise of mobile apps has changed the way software developers address users' needs. In recent years, the Requirements Engineering (RE) landscape has evolved significantly with the emergence of large-scale user feedback, especially app reviews, serving as a key source of actionable insights \cite{al2019app, guzman2014users, maalej2015toward, martin2017app, pagano2013user, tavakoli2018extracting, maalej2015bug}. However, popular apps generate large amounts of reviews, making manual analysis impractical \cite{pagano2013user}. Therefore, various tools have been developed over the past decade to aid this analysis \cite{tavakoli2018extracting}. 
Among several review-based RE tasks, accurately extracting requirements is a crucial step because the quality of the extracted requirements directly influences other essential tasks, such as clustering of requirements to organize similar requirements and sentiment analysis of requirements to identify the sentiment of user feedback \cite{de2021re}.


\textbf{Problem. }Recent research has highlighted several challenges in automated requirements extraction including (i) the inherent linguistic complexity of user-generated content, characterized by informal language, grammatical variations, and contextual nuances \cite{pagano2013user}; (ii) the multi-faceted nature of reviews, where a single text may contain opinions about multiple app aspects \cite{guzman2014users}; (iii) the significant noise and irrelevant information present in large review datasets \cite{johann2017safe}; and (iv) the domain-specific variations in terminology and user expectations \cite{dragoni2019unsupervised}. Efforts such as RE-BERT \cite{de2021re} and T-FREX \cite{motger2024t} have made notable progress by redefining requirements extraction as a Named Entity Recognition (NER) task, a specific type of token classification in which tokens referring to a particular entity type (e.g., dates, geopolitical entities, requirements, etc.) are labeled as such. RE-BERT leveraged context-dependent language models and cross-domain training to improve requirements extraction, while T-FREX experimented with different language models with a newly created crowdsourced user-annotated app reviews dataset. These methods represent advancements in moving beyond traditional rule-based techniques, which often struggle with semantic complexity and generalizability \cite{johann2017safe, guzman2014users, dragoni2019unsupervised}. However, these studies have still achieved a low F1 score in the range of 0.4-0.5, which shows the scope for further improvement and motivates this study.

\textbf{Solution. }In this study, we aim to improve requirements extraction by redefining it as an NER task based on a sequence-to-sequence (Seq2seq) generation problem instead of a traditional token classification problem. To achieve this, we propose a Seq2seq framework in which we first use GloVe word embeddings to transfer the words in reviews to low-dimensional vectors \cite{pennington-etal-2014-glove}. Next, we leverage a Bidirectional Long Short-Term Memory (BiLSTM) model \cite{schuster1997bidirectional} as an encoder to process the input sequence in two directions, as past and future information is equally important for the current output. Then, we adopt a self-attention mechanism \cite{vaswani2017attention} to help our model focus on the significant part, such as the named entity, in the input sequence. Next, we use a Long Short-Term Memory (LSTM) model \cite{hochreiter1997long} as a decoder to obtain the final hidden representations of the words in the input sequence. Finally, we use the condition random fields (CRF) model \cite{sutton2012introduction} to tag the sequence using the BIO (\textbf{B}eginning, \textbf{I}nside, and \textbf{O}utside) tags corresponding to the NER task. A CRF model is more sensitive to position logic \cite{lample2016neural}; hence, we do not directly use the decoder to generate the BIO tags.

\textbf{Rationale. }The Seq2seq framework is based on the encoder-decoder architecture that produces a target sequence of any length from a source sequence. In the context of the NER, it generates the sequence of BIO tags for the given input sequence of app reviews. The Seq2seq framework is widely used for various tasks like machine translation \cite{bahdanau2014neural} and, recently, for domain-specific NER, where it has shown remarkable performance \cite{hu2024tackling}. Thus, taking inspiration from these examples, in this study, we explore the utility of the Seq2seq framework for RE domain-specific NER to extract the requirements from app reviews. In our framework, we used GloVe embeddings to capture the semantic and syntactic information of the input because it is derived from large-scale corpora, resulting in semantically richer and more stable representations compared to other embeddings \cite{pennington2014glove}. Next, we chose LSTMs as an encoder and decoder in our framework, which has been a standard choice for similar problems \cite{zhu2020fine}; however, we used a BiLSTM as an encoder to process the input sequence, considering both future and past information to get better results. Furthermore, we integrated the self-attention mechanism and a CRF model to focus on contextually relevant elements and generate coherent and accurate outputs.

Our study excluded GPT models that use decoder-only architectures due to their inherent limitations in performing domain-specific NER tasks \cite{rafique2024decoding}. GPT has been extensively employed across various RE tasks, such as requirements classification \cite{arora2024advancing} and requirements elicitation \cite{ronanki2023investigating}, demonstrating advantages in terms of increased productivity and cost efficiency \cite{marques2024using}. However, in the context of the domain-specific NER task, zero-shot or few-shot GPT has exhibited subpar performance and has been computationally expensive compared to the Seq2seq and traditional NER methods that utilize fine-tuning on the labeled datasets \cite{rafique2024decoding, gupta2023comparison}.


\textbf{Contributions. }We propose a novel reformulation of the requirements extraction task as an NER task utilizing the sequence-to-sequence generation approach and introduce a Seq2seq framework that integrates a BiLSTM encoder and a LSTM decoder, complemented by the self-attention mechanism, GloVe embeddings, and a CRF model for enhanced performance. We conduct a comparative analysis of our framework against existing state-of-the-art models, demonstrating performance on par with RE-BERT while significantly surpassing T-FREX. To our knowledge, this study is the first to apply the proposed set of techniques for requirements extraction from app reviews. We have made our source code publicly available\footnote{\url{https://doi.org/10.5281/zenodo.15185900}}.


\textbf{Structure. }In Section \ref{prem} and Section \ref{back}, we outline the preliminaries and discuss the related works, respectively. Section \ref{approach} presents the proposed approach, and Section \ref{method} outlines the methodology, including research questions (RQs) and datasets. Section \ref{results} reports and discusses preliminary results, and Section \ref{limit} presents the limitations. Finally, Section \ref{conclusion} presents our research plan and offers concluding remarks.

\section{Preliminaries} \label{prem}


\textbf{Named Entity Recognition (NER).} NER is one of the most common token classification tasks in the context of natural language understanding \cite{nasar2021named}. It identifies specific entities, such as individuals, locations, and organizations, within a given text using the BIO tags \cite{nasar2021named}. In the domain of requirement engineering and specifically for requirements extraction tasks, NER is used to identify requirements from the natural language texts using BIO tags, where \textbf{B} refers to the beginning of a requirement, \textbf{I} means the continuation of a requirement, and \textbf{O} means irrelevant information \cite{sorathiya2024towards}. Additionally, NER based on the Seq2seq generation approach aims to generate a sequence of BIO tags for the given input sequence \cite{zhu2020fine}. \\

\noindent\textbf{Requirements Extraction.} In the landscape of mobile apps, the functional requirement refers to a property, feature, or characteristic of an app \cite{motger2024t}. Previous studies have used feature or requirement terms to extract functional requirements from app reviews \cite{motger2024t} \cite{de2021re}. Similarly, in this study, we use requirement to refer to a functional requirement or a feature in app reviews. The requirements extraction task refers to the automatic and accurate identification of the requirements included in the natural language text, which is unstructured, potentially noisy, and contains varied vocabulary \cite{johann2017safe}. For example, in the review ``Can you add audio format for text to speech?", \textit{audio format} and \textit{text to speech} are extracted as requirements. In another example review, ``I also like the 'rewind' button.", \textit{rewind button} is extracted as a requirement. 

\section{Related Work} \label{back}
An automatic app review analysis pipeline consists of multiple stages, such as requirement extraction, classification, clustering, and sentiment analysis. However, while other stages may be included, our focus will remain on the extraction stage when discussing related works.

GuMa \cite{guzman2014users}, SAFE \cite{johann2017safe}, and ReUS \cite{dragoni2019unsupervised} proposed linguistic rules-based information extraction strategies to extract requirements from app reviews. GuMa employed a collocation finding algorithm from the NLTK toolkit \cite{bird2006nltk} for requirements extraction. SAFE extracted requirements through 18 part-of-speech patterns and 5 sentence patterns. ReUS investigated linguistic rules based on grammatical class patterns and semantic dependencies by analyzing aspects to extract requirements. However, these rule-based approaches show limitations because textual expressions can be represented by rules with high complexity, a large number of rules can generate conflicts, and rules extracted by analyzing data from a specific domain generally do not meet the context of other domains \cite{aggarwal2018machine}.

RE-BERT \cite{de2021re} transformed the requirements extraction task into an NER task based on the classification of BIO tags using an encoder-only BERT language model. It leveraged the ``Local Context" attention mechanism that emphasized tokens situated near potential requirements while simultaneously accounting for global contextual information. Furthermore, the BIO tag classification technique offered granularity and significantly enhanced the precision and recall of requirement extraction. Additionally, it employed a cross-domain training strategy, harnessing manually curated datasets from various domains to augment its generalization capabilities, thereby facilitating effective extraction in previously unseen app domains. This methodology yielded superior performance to traditional approaches: SAFE, GuMa, and ReUS.

T-FREX \cite{motger2024t} also utilized the NER task based on the classification of BIO tags to extract requirements from app reviews. It employed various pre-trained Transformer-based encoder-only language models (BERT, RoBERTa, XLNet) that were fine-tuned on app reviews annotated with verified ground-truth requirements. It leveraged crowdsourced annotations to develop a comprehensive set of requirements and then labeled them within app reviews, resulting in a large annotated dataset of app reviews. Furthermore, it adopted a cross-domain training strategy to improve the domain adaptability and efficiently identify requirements across various app domains. T-FREX also integrated meticulous evaluation against baseline methodologies, consistently surpassing the traditional syntactic method, SAFE, particularly in the identification of undocumented requirements. External human evaluations further substantiated its efficacy, highlighting its value in extracting novel requirements from unseen datasets.

Both RE-BERT and T-FREX showed improvement over traditional methods, SAFE, GuMa, and ReUS; however, they still reported a low F1 score in the range of 0.4-0.5. These suboptimal values indicate significant potential for further enhancements and serve as a driving force for this research. Furthermore, we implemented the NER technique akin to RE-BERT and T-FREX, but we adopted a Seq2seq framework rather than the traditional classification approach. The Seq2seq framework has been widely used for various Natural Language Processing (NLP) tasks, such as machine translation \cite{bahdanau2014neural}, text summarization \cite{nallapati2016abstractive}, and chatbots \cite{vinyals2015neural}. Recently, it has been used for domain-specific NER and has shown superior performance over traditional NER methods due to its ability to fully comprehend the input sentence using the encoder-decoder architecture \cite{zhu2020fine, hu2024tackling}.  Despite such promising results for various NLP tasks, the Seq2seq framework has not yet been leveraged in the RE domain to extract requirements from app reviews, which we address in this study.

\section{Proposed Seq2seq Approach} \label{approach}
\begin{figure}[h]
    \centering
    \includegraphics[width=1\linewidth]{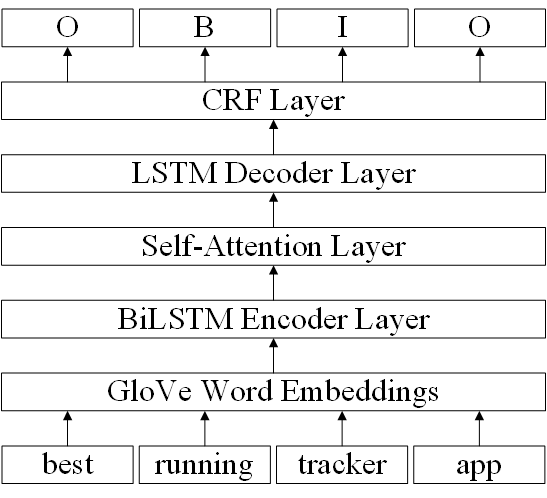}
    \caption{Our Seq2seq framework is built using the GloVe embeddings, BiLSTM encoder, self-attention mechanism, LSTM decoder, and a CRF model.}
    \label{fig:model}
\end{figure}

Figure \ref{fig:model} illustrates our Seq2seq framework, from which it can be observed that it is constructed using the GloVe embeddings layer, BiLSTM encoder layer, self-attention layer, LSTM decoder layer, and a CRF layer.

\textbf{Embedding Layer}: The embedding layer transforms words into dense vector representations using GloVe embeddings. We convert each word in the input sentence to a one-hot vector, which is then mapped to a low-dimensional vector from a pre-trained embedding lookup table. This process converts a sequence of words into dense vector representations that capture semantic relationships between words, preparing the input for further processing by the model's encoder layer.

\textbf{BiLSTM Encoder Layer}: To encode the input sequence, the future information and past information are equally important for the current output; thus, we adopt the BiLSTM model to process the future and past elements simultaneously. A BiLSTM model is composed of two parallel LSTM models, that is, the forward LSTM and the backward LSTM. In the forward LSTM, the current state is calculated by employing input, forget, and output gates that control the flow of information, considering both the current input and the previous state. Conversely, the backward LSTM processes the sequence in the reverse direction, thus capturing future context. By concatenating the hidden states obtained from both models, we create a comprehensive representation that synthesizes contextual information from both directions, thereby enabling a more nuanced understanding of each input element.

\textbf{Self-Attention Layer}: The self-attention layer helps the model focus on the most significant parts of the input sequence by dynamically assigning different weights to each element. Using an attention mechanism, the model can effectively handle long-term dependencies, giving more importance to contextually relevant elements. The self-attention process involves computing attention scores, applying a softmax to generate attention weights, and then creating a weighted representation of the input sequence that captures the most important information for the current output.

\textbf{Decoder Layer}: The decoder layer uses a single-layer LSTM to generate the final tag representations. Its input is complex, combining the encoder's hidden state after self-attention, the previously generated tag representation, and the previous hidden state. The decoder uses input, forget, and output gates to control the information flow, processing these inputs to generate a sequence of tag representations. Through this process, the model progressively refines and predicts the most appropriate tag for each token in the sequence.

\textbf{CRF Layer}: Traditional Seq2seq models utilize the softmax classifier to generate the tag, however, it is likely to generate mistaken entities, which is illogical. For example, it is possible to generate the tag for the requirement `GPS tracking’ with `GPS’ as `B’ and `tracking’ as `B’, which is incorrect. To address this, the CRF layer adds constraints to ensure more accurate tag generation. The CRF considers both the individual tag probabilities for each word and the transition probabilities between tags, effectively preventing impossible or nonsensical tag sequences. By evaluating the entire sequence's tag transitions and probabilities, the CRF selects the most probable and logically consistent tag sequence for the input sentence \cite{panchendrarajan2018bidirectional}.

\section{Methodology} \label{method}

\begin{figure*}[htb]
    \centering
    \includegraphics[width=1\linewidth]{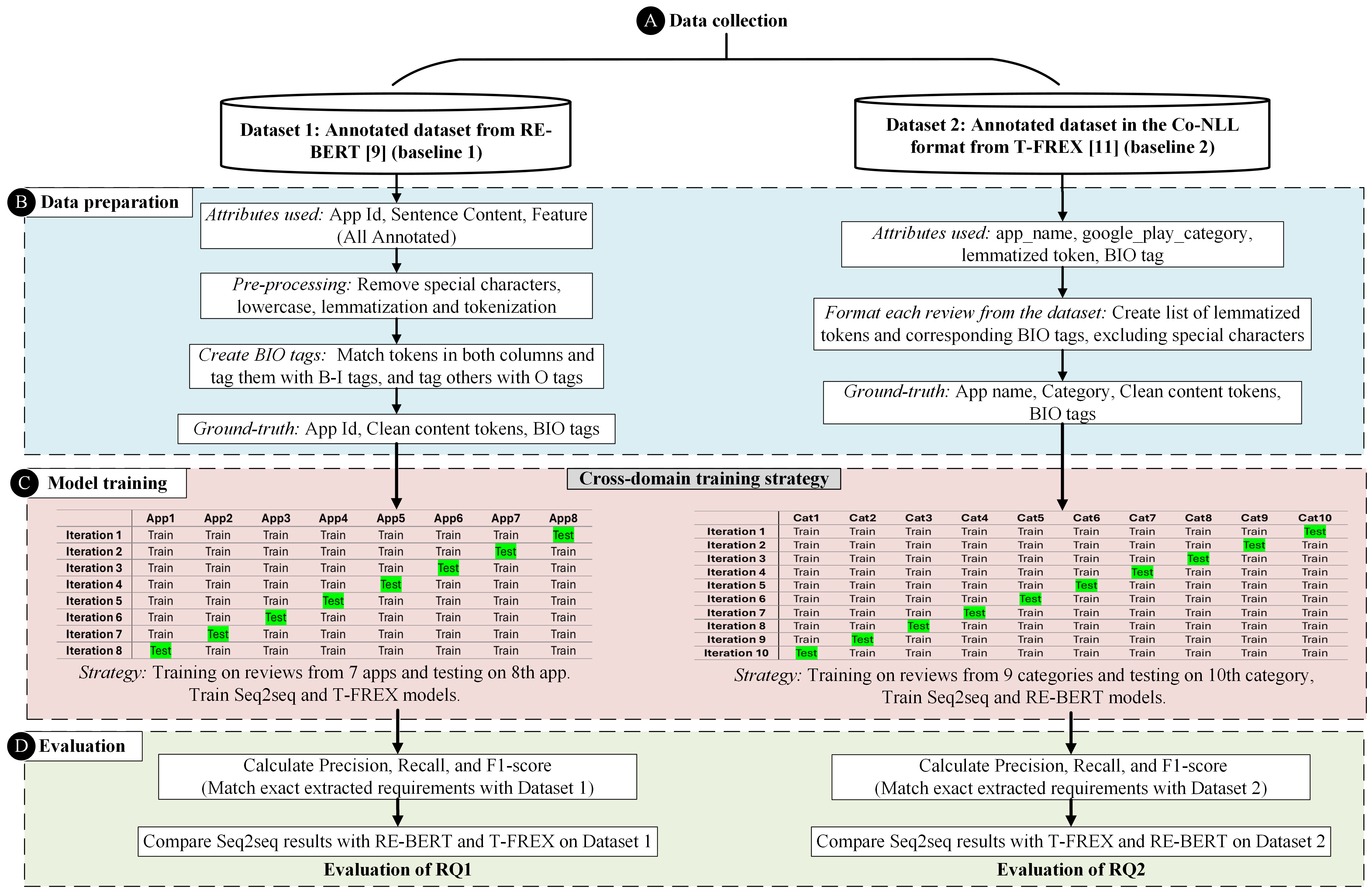}
    \caption{Research design overview. \textbf{A}: Data collection from two baselines: RE-BERT and T-FREX. \textbf{B}: Preprocessing pipelines to clean the data and convert it to the required format of BIO tags. \textbf{C}: Training of the Seq2seq framework using the cross-domain training strategy employed by the baselines, where a set of apps or categories are selected for training and the remaining are used for evaluation. \textbf{D}: Evaluation of Seq2seq framework in alignment with our RQs.}
    \label{fig:design}
\end{figure*}

We demonstrate the hypothesis that the Seq2seq framework based on the encoder-decoder architecture can significantly enhance requirements extraction by redefining this process as an NER task based on the sequence-to-sequence generation approach. Thus, to assess the validity of our Seq2seq framework, we guide our research through the following RQs:\\
\textbf{RQ1:} \textit{What is the effectiveness of the Seq2seq framework compared to the baseline RE-BERT and T-FREX for requirements internally annotated by human coders? }\\ To answer this question, we utilized the dataset of app reviews from RE-BERT, which is annotated by human coders. \\ \\
\textbf{RQ2:} \textit{What is the effectiveness of the Seq2seq framework compared to the baseline RE-BERT and T-FREX for crowdsourced user-annotated requirements? } \\To answer this question, we utilized the dataset of app reviews from T-FREX, which is labeled with crowdsourced requirements.

Our RQs assess the quality of the proposed Seq2seq framework on two separately curated datasets and compare it with the state-of-the-art for requirements extraction from app reviews. Figure \ref{fig:design} shows an overview of our research methodology, and we elaborate on the details below.

\subsection{Data collection. } We utilized two ground-truth datasets for the evaluation of our proposed approach and compared it with the two baseline studies. Dataset 1 is collected from RE-BERT, which consists of 1,000 manually labeled reviews, and Dataset 2 is collected from T-FREX, which consists of 23,816 crowdsourced annotated reviews. 

\begin{table*}[h]
    \centering
    \renewcommand{\arraystretch}{1.2}
    \caption{Dataset 1 statistics from RE-BERT \cite{de2021re}. It contains 1,000 reviews manually labeled with 1,172 unique requirements.}
    \label{tab:d1}
    \begin{tabular}{lrrrrR{2cm}R{2cm}R{2cm}}
        \textbf{App} & \textbf{\#reviews} & \textbf{\#labeled reviews} & \textbf{\#sentences} & \textbf{\#requirements} & \textbf{\#distinct requirements} & \textbf{\#single-word requirements} & \textbf{\#multi-word requirements}\\
        \hline
        eBay & 1,962 & 125 & 294 & 206 & 167 & 78 & 128\\
        Evernote & 4,832 & 125 & 367 & 295 & 259 & 82 & 213\\
        Facebook & 8,293 & 125 & 327 & 242 & 204 & 80 & 162\\
        Netflix & 14,310 & 125 & 341 & 262 & 201 & 94 & 168\\
        Photo editor & 7,690 & 125 & 154 & 96 & 80 & 39 & 57\\
        Spotify & 14,487 & 125 & 227 & 180 & 145 & 69 & 111\\
        Twitter & 63,628 & 125 & 183 & 122 & 99 & 39 & 83\\
        WhatsApp & 248,641 & 125 & 169 & 118 & 100 & 49 & 69\\
        \hline
        \textbf{Total} & 341,843 & 1,000 & 2,062 & 1,521 & 1,172 & 530 & 991
    \end{tabular}
\end{table*}
\begin{table*}[h]
    \centering
    \caption{Dataset 2 statistics from T-FREX \cite{motger2024t}. It contains 23,816 reviews annotated with 198 unique crowdsourced requirements.}
    \label{tab:d2}
    \begin{tabular}{p{2cm}rR{1.5cm}R{1.5cm}rR{2cm}R{2cm}R{2cm}}
        \textbf{App Category} & \textbf{\#apps} & \textbf{\#reviews} & \textbf{\#sentences} & \textbf{\#requirements} & \textbf{\#distinct requirements} & \textbf{\#single-word requirements} & \textbf{\#multi-word requirements}\\
        \hline
        Productivity & 137 & 7,348 & 8,604 & 10,141 & 77 & 8,542 & 1,599 \\
        Communication & 51 & 7,003 & 8,135 & 10,595 & 54 & 9,419  & 1,176 \\
        Tools & 58 & 4,321 & 5,402 & 6,773 & 50 & 6,384 & 389 \\
        Social & 14 & 819 & 899 & 1,115 & 26 & 1,030 & 85\\
        Health & 75 & 2,154 & 2,330 & 2,691 & 23 & 2,572 & 119 \\
        Personalization & 6 & 112 & 118 & 124 & 19 & 117 & 7\\
        Travel & 19 & 530 & 602 & 704 & 17 & 671 & 33\\
        Maps & 31 & 284 & 315 & 352 & 12 & 336 & 16 \\
        Lifestyle & 12 & 344 & 391 & 459 & 10 & 368 & 91 \\
        Weather & 65 & 901 & 984 & 1,105 & 7 & 1,062 & 43 \\
        \hline
        \textbf{Total} & 468 & 23,816 & 27,780 & 34,059 & 198 & 30,501 & 3,558
    \end{tabular}
\end{table*}

Table \ref{tab:d1} shows statistics about Dataset 1. The authors in \cite{de2021re} indicated that the dataset was produced through a thorough manual annotation procedure carried out by two individuals, adhering to established guidelines from the existing literature regarding data selection, labeling, agreement assessment, and reliability evaluation. Initially, a total of 341,843 reviews were collected in unrestricted English text submitted by users of apps available on the Google Play and Amazon App stores. These reviews originated from 8 apps spanning various categories and time frames. A random selection of these reviews was manually examined by the annotators. Ultimately, 1,172 unique requirements were identified from 1,000 reviews, organized into 2,062 sentences. Within the compiled list of annotated requirements, 530 requirements consisted of a single word, while 991 comprised multiple words.

Table \ref{tab:d2} shows statistics about Dataset 2. The authors in \cite{motger2024t} indicated that the dataset originated from an extended version of a previously published collection of 639 mobile apps with 622,370 reviews \cite{motger2023mobile}, with several critical methodological enhancements. The researchers focused on the 10 most frequent Google Play categories and deliberately excluded game-related categories, resulting in a dataset of 23,816 reviews across 468 apps and 10 categories. To create the ground-truth dataset, they used the \textit{AlternativeTo} platform as a primary source to get crowdsourced requirement annotations. Next, for requirement labeling in the reviews, the dataset was processed using Stanza's neural pipeline \cite{qi2020stanza} to transform the corpus into a standardized CoNLL-U format, which included comprehensive linguistic annotations such as tokenization, part-of-speech tagging, morphological requirement extraction, and lemmatization. Then, the previously collected crowdsourced requirements were precisely transferred into the reviews, where each token in the review was annotated with labels indicating whether it represented an outside entity (O), the beginning of a requirement (B-feature), or a continuation of a requirement (I-feature) entity. As a result, 198 distinct requirements were annotated in 27,780 sentences. 

\subsection{Data preparation. }From Dataset 1, we selected the \textit{App Id}, \textit{Sentence Content}, and \textit{Feature (All Annotated)} columns where \textit{App Id} represents the id of the app for the corresponding review, \textit{Sentence Content} represents a single sentence of the review, and \textit{Feature (All Annotated)} represents all the requirements extracted from the corresponding review sentence. Then we processed the data of \textit{Sentence Content} and \textit{Feature (All Annotated)} columns by applying the following steps: removal of special characters, lowercasing, lemmatization, and tokenization. Finally, we matched the tokens of the \textit{Feature (All Annotated)} column with the tokens of the \textit{Sentence content} column and tagged the matching tokens with B and I tags to represent the requirements. The rest of the tokens were tagged with the O tag. As a result, our ground-truth Dataset 1 contained three columns: \textit{App Id}, \textit{Clean content tokens} (processed review tokens), and \textit{BIO tags} (list of BIO tags).

From Dataset 2, containing CoNLL-U documents, we selected \textit{app\_name}, \textit{google\_play\_category}, \textit{lemmatized\_token}, and \textit{BIO\_tag}, where \textit{app\_name} is the app name for the corresponding review, \textit{google\_play\_category} is the category of an app on Google Play store, \textit{lemmatized\_token} is the lemmatized form of the review token, and \textit{BIO\_tag} is the tag for the corresponding review token. Next, for each document representing the CoNLL-U form of each review, we created a list of lemmatized tokens and corresponding BIO tags by excluding special characters. Additionally, we changed the B-feature and I-feature tags to B and I to represent the requirements and unify our modeling approach across both datasets. As a result, our ground-truth Dataset 2 contained four columns: \textit{App name}, \textit{Category}, \textit{Clean content tokens} (lemmatized review tokens), and \textit{BIO tags} (list of BIO tags).

\subsection{Model training. }Similar to the two baselines, to train and evaluate our proposed Seq2seq framework, we performed a cross-validation process based on multiple app domains/categories. This strategy is referred to as a cross-domain training strategy, in which we select a set of app domains/categories for training and evaluate the model on the remaining app domain/category. This strategy evaluates the proposed framework to extract requirements from app domains/categories with no labeled data, since the manual labeling process is an expensive task. Furthermore, incorporating different domains during the training stage is a robust approach for learning a model with better generalization capabilities \cite{de2021re}.

For Dataset 1, the model was trained on the labeled data with seven different apps and evaluated for extracting requirements from an eighth app. This process was repeated until all the apps were used as test data. Similarly, for Dataset 2, the training process was applied from the labeled data with nine different categories, and then the model was evaluated on the tenth category. This process was repeated until all categories were used as test data. Additionally, we performed 15 runs for each app/category to evaluate the stability of our framework and reported the average result across all the runs.

Next, to transform the data into a format directly compatible with our model's input requirements, we applied the following steps. First, we created a comprehensive vocabulary, mapping each unique word in the dataset to a distinct index, and then transformed the input data into numerical index representations using this vocabulary, ensuring consistent encoding across the dataset. To maintain uniform input dimensions for the model processing, we padded the input sequences to the maximum length in the dataset. Finally, we converted these input sequences into PyTorch tensors. For model training, we used a learning rate of 0.001, a word embedding dimension of 300, a batch size of 32, the Adam optimization algorithm, and Negative log-likelihood as a loss function.

\subsection{Evaluation design. }To evaluate the performance of our framework, we used a requirement-level evaluation method by computing the quality metrics for requirement extraction. We first identified the predicted BIO class of each token in the review, and then we extracted all tokens classified as \textit{B} or \textit{I} as a requirement. When two or more consecutive tokens in the review were classified as \textit{B} or \textit{I}, we unified these tokens to be considered as a single extracted requirement. Next, we compared these extracted requirements with the ground-truth dataset using Precision (P), Recall (R), and the F1 score as evaluation metrics. P refers to the percentage of the requirements extracted that are correct, and R refers to the percentage of requirements in the ground-truth dataset that the model successfully extracted. The F1 score (\(F1 = \frac{2*P*R}{P+R}\)) is the weighted harmonic average of P and R, which represents the performance of the model in a comprehensive way.


Next, we structured the evaluation results in alignment with our RQs and cross-domain training strategy. For RQ1, we reported the F1 score for each of the eight apps in Dataset 1; for RQ2, we reported the F1 score for each of the ten categories in Dataset 2. We compared these results with the baselines, RE-BERT and T-FREX. Note that T-FREX used three models in their experiments; however, we chose the best-performing model, i.e., XLNet, to compare with our approach.

\subsection{Computational resources. }
The experiments were conducted on an NVIDIA GeForce RTX 4090 GPU of 40 GB of RAM, NVIDIA-SMI driver version 546.09, and a 24-core CPU setup. We implemented our framework using Python 3.12 with CUDA version 11.8 and PyTorch version 2.3.1+cu118. We used NumPy and Pandas for linear algebra operations.

\section{Preliminary Results and Discussion} \label{results}

\begin{table}[t]
    \centering
    \renewcommand{\arraystretch}{1.2}
    \caption{RQ1: Comparison of F1 score of Seq2seq with RE-BERT and T-FREX on Dataset 1. Seq2seq performed similarly to RE-BERT and significantly outperformed T-FREX.}
    \label{tab:r1}
    \begin{tabular}{l|EEE}
        \textbf{App} & \multicolumn{1}{r}{\textbf{Seq2seq}} & \multicolumn{1}{r}{\textbf{RE-BERT}} & \multicolumn{1}{r}{\textbf{T-FREX}} \\ \\
        eBay & 0.45 & 0.40 & \multicolumn{1}{r}{0.09} \\
        Evernote & 0.41 & 0.45 & \multicolumn{1}{r}{0.10} \\
        Facebook & 0.42 & 0.43 & \multicolumn{1}{r}{0.11} \\
        Netflix & 0.41 & 0.47 & \multicolumn{1}{r}{0.13} \\
        Phone editor & 0.58 & 0.66 & \multicolumn{1}{r}{0.15} \\
        Spotify & 0.46 & 0.44 & \multicolumn{1}{r}{0.11} \\
        Twitter & 0.48 & 0.57 & \multicolumn{1}{r}{0.14} \\
        WhatsApp & 0.51 & 0.48 & \multicolumn{1}{r}{0.13} \\
        \hline
        \textbf{Mean} & \multicolumn{1}{r}{0.47} & \multicolumn{1}{r}{0.48} & \multicolumn{1}{r}{0.12}
    \end{tabular}
\end{table}

\begin{table}[t]
    \centering
    \renewcommand{\arraystretch}{1.2}
    \caption{RQ2: Comparison of F1 score of Seq2seq with RE-BERT and T-FREX on Dataset 2. Seq2seq significantly outperformed RE-BERT and T-FREX.}
    \label{tab:r2}
    \begin{tabular}{l|FFF}
        \textbf{Category} & \multicolumn{1}{l}{\textbf{Seq2seq}} & \multicolumn{1}{l}{\textbf{RE-BERT}} & \multicolumn{1}{l}{\textbf{T-FREX}} \\ \\
        Productivity & 0.94 & 0.85 & \multicolumn{1}{r}{0.30} \\
        Communication & 0.95 & 0.88 & \multicolumn{1}{r}{0.22} \\
        Tools & 0.97 & 0.88 & \multicolumn{1}{r}{0.23} \\
        Social & 0.98 & 0.90 & \multicolumn{1}{r}{0.57} \\
        Health & 0.96 & 0.88 & \multicolumn{1}{r}{0.44} \\
        Personalization & 0.98 & 0.91 & 0.86 \\
        Travel & 0.98 & 0.89 & \multicolumn{1}{r}{0.56} \\
        Maps & 0.95 & 0.88 & \multicolumn{1}{r}{0.19} \\
        Lifestyle & 0.97 & 0.89 & \multicolumn{1}{r}{0.60} \\
        Weather & 0.96 & 0.90 & \multicolumn{1}{r}{0.46} \\
        \hline
        \textbf{Mean} & \multicolumn{1}{r}{0.96} & \multicolumn{1}{r}{0.88} & \multicolumn{1}{r}{0.44}
    \end{tabular}
\end{table}



\textbf{Answering RQ1. }Table \ref{tab:r1} presents a comparative analysis of F1 scores between Seq2seq and baseline models across eight different apps in Dataset 1. Seq2seq and RE-BERT demonstrated remarkably similar performance, significantly outperforming T-FREX, with a marginal difference in their mean F1 scores: Seq2seq achieved an F1 score of 0.47, while RE-BERT slightly edged ahead with an F1 score of 0.48. When examining the performance of Seq2seq and RE-BERT across individual apps, nuanced variations were observed. Seq2seq demonstrated superior performance in three apps: eBay, Spotify, and WhatsApp. Conversely, RE-BERT outperformed in four apps: Evernote, Netflix, Phone Editor, and Twitter. Facebook showed almost identical performance between the two models, with a negligible 0.01 difference. T-FREX achieved the lowest F1 score across all the apps with an average of just 0.12. \\

\textbf{Answering RQ2. }Table \ref{tab:r2} presents a comparative analysis of F1 scores between Seq2seq and baseline models across ten categories in Dataset 2. The results show that Seq2seq achieved the highest average F1 score of 0.96, compared to RE-BERT with an average F1 score of 0.88, and significantly outperformed T-FREX with an average F1 score of 0.44. Looking at individual category performances of Seq2seq and T-FREX, Seq2seq demonstrated remarkable consistency, maintaining the F1 score between 0.94 and 0.98 across all categories. In contrast, T-FREX showed extreme variability, with the F1 score ranging from a low of 0.19 (Maps) to a high of 0.86 (Personalization). The most notable performance gaps were in the Communication (Seq2seq: 0.98 vs. T-FREX: 0.22), Tools (Seq2seq: 0.97 vs. T-FREX: 0.23), and Maps (Seq2seq: 0.97 vs. T-FREX: 0.19) categories. RE-BERT also showed improvement over T-FREX across all categories, with an F1 score ranging from 0.85 to 0.91. \\

\begin{figure}
    \centering
    \includegraphics[width=1\linewidth]{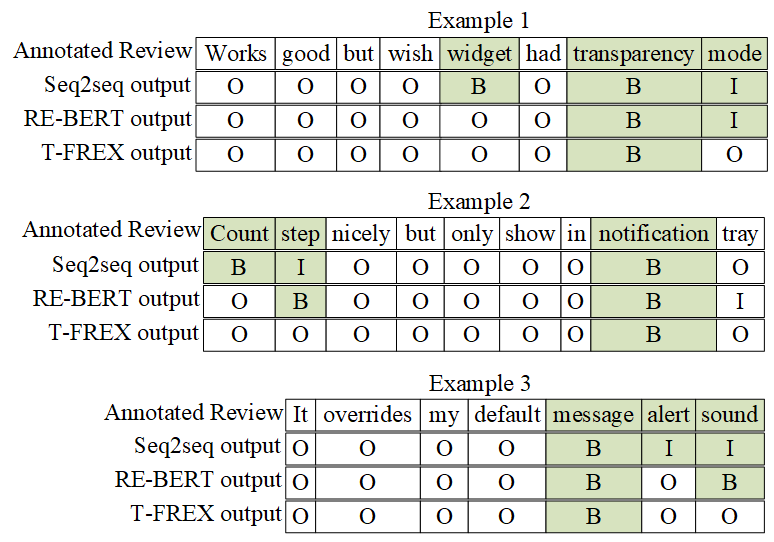}
    \caption{Requirements extraction examples comparing the Seq2seq framework with RE-BERT and T-FREX. Annotated Review represents the original review from the ground-truth dataset, annotated with requirements, highlighted in green. In the output, B and I tokens representing the requirements in the annotated review are highlighted green.}
    \label{fig:result}
\end{figure}

\textbf{Discussion of Results. }
Figure \ref{fig:result} shows an example of requirements extracted using the proposed Seq2seq framework. Our preliminary results show the potential of leveraging the Seq2seq framework for the efficient requirements extraction task by comparing it with the state-of-the-art RE-BERT and T-FREX. Seq2seq achieved an average F1 score of 0.47 on Dataset 1 and 0.96 on Dataset 2. We believe that this notable difference in the model's performance across both datasets is due to the methodologies that were employed to curate them. Dataset 1 was internally annotated by human coders, very limited in the number of apps (8 apps), and focused on productivity and communication domains, excluding more expert domains like navigation, sports, or weather. However, Dataset 2 was created using crowdsourced annotations from real users for 468 apps from 10 different app domains.

Despite utilizing a highly sophisticated transformer-based model, which is renowned for its capabilities, T-FREX demonstrated below-average performance (metrics) for both datasets. We note that this limitation of T-FREX could be due to the omission of supplementary attention mechanisms such as those utilized in Seq2seq and RE-BERT, which subsequently results in its lack of ability to effectively discern and identify contextually relevant components present within the input text.

Seq2seq significantly outperformed RE-BERT on Dataset 2 and demonstrated similar performance for Dataset 1; however, we emphasize that RE-BERT used substantial computational resources during the model's training and evaluation process. For Dataset 2, the RE-BERT model training and evaluation process required an extensive duration of 50 minutes for each iteration, whereas, Seq2seq was able to complete its training and evaluation in a significantly shorter timeframe of just 15 minutes. We attribute this stark contrast in time efficiency to a considerable advancement associated with Seq2seq's framework, which highlights its significant potential for practical application in environments characterized by limited resources, especially when dealing with large datasets.


\section{Limitation and Threats to Validity} \label{limit}
\textbf{Construct threats.} The potential threat to the construct validity of our study pertains to the suitability of app review datasets. Dataset development is labor-intensive and susceptible to reader bias. Therefore, to address this concern, we selected two different datasets curated using two different techniques: first, manual labeling by a small group of annotators \cite{de2021re} and second, crowdsourced annotations \cite{motger2024t}. Another potential threat comes with the choice of evaluation metrics; therefore, we adopted P, R, and F1 metrics in this study, which are widely recognized and recommended for assessing such tasks within the SE domain.

\textbf{Internal threats. }The potential threats to the internal validity of our study mainly relate to the formalization of \textit{requirement} definition and the choice of hyperparameters to train the Seq2seq model. To address these concerns, we referred to the related works \cite{de2021re, motger2024t} in the domain of requirements extraction and formalized the definition of \textit{requirement} to represent functional requirements or features in app reviews. Furthermore, we selected the standard values of hyperparameters in our Seq2seq model that have been used in previous works \cite{zhu2020fine}; however, we acknowledge that altering these values could affect the results.


\section{Research Plan and Conclusion} \label{conclusion}
In this work, we proposed the reformulation of the requirements extraction task as an NER task using the sequence-to-sequence generation approach. We presented a Seq2seq framework consisting of a BiLSTM encoder and LSTM decoder, enhanced with a self-attention mechanism, GloVe embeddings, and a CRF model, to identify and extract requirements from app reviews effectively. Through experiments on two different datasets, we demonstrated that our framework outperformed existing solutions, achieving a mean F1 score of 0.47 on Dataset 1 and 0.96 on Dataset 2. We present our research plan below to further assess and improve our framework.

\begin{itemize}[leftmargin=*]
    \item We used GloVe embeddings to represent our input text into vectors; however, we will examine the software domain-specific word embeddings to improve performance. We will also conduct an ablation study to identify the impact of CRF in our Seq2seq framework. We will also assess transformer-based encoder-decoder models, such as T5, and integrate evaluation metrics, such as accuracy or ROC-AUC, to provide a more comprehensive analysis of our framework. Additionally, we will perform a sensitivity analysis to identify the performance sensitivity of our model for different sequence lengths and padding. We will also employ additional measures such as memory, energy, cost, and computation to evaluate the resource efficiency of the framework. \\
    \item We will integrate the confidence score with the extracted requirements in our framework by adding a confidence estimation layer \cite{li2021confidence} to assess the quality of the extracted requirements and determine the actionable requirements based on the confidence level.  Further, we plan to transform the extracted natural language textual requirements from our model into fine-grained requirements, which can be easily interpreted and addressed. Additionally, we plan to generate themes for the fine-grained requirements that will help to analyze the user preferences at scale. \\
    \item Sentiment analysis in app reviews has helped in identifying users' likes or/and dislikes about an app \cite{guzman2014users}, understanding how the app is perceived among users \cite{fu2013people}, and prioritizing bug fixes based on negative sentiments \cite{chen2021should}. Therefore, we also plan to integrate the user sentiment score as an output of our framework, which might enable app developers to prioritize requirements based on user preferences. \\
    \item As mobile applications become increasingly integral to our daily lives, concerns about ethics have grown drastically \cite{sorathiya2024ethical}. Recent research by Sorathiya et al. \cite{sorathiya2024towards} and Kara{\c{c}}am et al. \cite{karaccam2024uncovering} has shown significant potential for leveraging app reviews to address users' ethical concerns. Thus, we envision extending our approach to extract ethical concerns-related requirements to resolve users' ethical concerns, such as privacy, fairness, or transparency.
\end{itemize}

\section*{Acknowledgement}
This research is supported and funded by the NSERC Alliance-Alberta Innovates Advance Program Stream I program. Also, we thank the anonymous reviewers for their constructive feedback, which has helped improve this work further.
\bibliographystyle{IEEEtran}
\bibliography{ref}

\end{document}